\documentclass{ptapap}
\usepackage[]{natbib}

\author{Joyce A. Guzik}[LANL]
\author{Chris Fryer}[LANL]
\author{Christopher J. Fontes}[LANL]
\affil[LANL]{Los Alamos National Laboratory, 
 Los Alamos, NM 87545, USA}
\title{Opacity Effects on Pulsations \\of A-Type Stars}

\begin{document}

\maketitle

\begin{abstract}

The effects of opacities in driving pulsations and extending the pulsation instability region for B-type stars (SPB and $\beta$ Cep variables) have recently been explored \citep[see, e.g.,][]{2015A&A...580L...9W, 2017arXiv170101258W, 2017MNRAS.466.2284D, 2016MNRAS.455L..67M, 2014A&A...565A..76C}.  For these stars, the pulsations are driven by the opacity bump of Fe-group elements at $\sim$200,000 K in the stellar envelope.  For $\delta$ Scuti pulsations in A-type  stars, pulsations are driven instead by the $\kappa$ effect in the second helium ionization region at $\sim$50,000 K.  Opacity modifications investigated for B-type stars may also have an effect on the structure of A-type stars and the instability region.  Studies of the pulsation instability region for A-type stars show that other factors also influence the pulsations, for example, helium abundance and diffusive settling.  Here we explore some of these effects, focusing on modeling of A-type stars.  We compare results using the new LANL OPLIB \citep{2016ApJ...817..116C}  vs. LLNL OPAL \citep{1996ApJ...464..943I} opacities for the \cite{2009ARA&A..47..481A} solar mixture.  We comment on prospects for using BRITE observations of A-type stars to constrain opacities.

\end{abstract}

\section{Introduction}
Opacities, element abundances, and convection affect pulsation instabilities in nearly all types of stars. Studies of pulsating B-type stars have motivated opacity investigations.  SPB/$\beta$ Cep stars show fewer modes than expected, and hybrids that are not predicted \cite[see, e.g.,][re.~$\nu$ Eri and 12 Lac]{2008MNRAS.385.2061D}.  Increasing the Fe/Ni opacities by a factor of 1.75, as inferred from Sandia pulsed-power experimental data \citep{2015Natur.517...56B}, widens the hybrid instability region enough to include 12 Lac \citep{2016MNRAS.455L..67M}.  Opacity modifications improve the match to pulsation observations for $\nu$ Eri \citep{2017MNRAS.466.2284D}.  The SPB and $\beta$ Cep pulsation-instability regions are somewhat wider \citep{2015A&A...580L...9W} using the new Los Alamos OPLIB \citep{2016ApJ...817..116C} opacities.

The ``solar abundance problem'' also motivates opacity increases for solar models.  Recent standard solar models taking into account uncertainties in input physics still show disagreement between the sound speed inferred from solar $p$ modes and that of the standard model using the latest abundance determinations \citep{2017ApJ...835..202V}.  The sound-speed discrepancy becomes slightly smaller using the new LANL OPLIB opacities instead of the LLNL OPAL opacities \citep{2016ApJ...817..116C, 2016IAUFM..29B.532G}.

Discrepancies with observations motivate opacity modifications for B-type stars and for the Sun.  Here we explore the consequences of opacity modifications for A-type stars.

\section{Stellar Evolution and Pulsation Models}

We use as our testbed for opacity studies the evolution of 2 M$_\odot$ models with OPAL \citep{1996ApJ...464..943I} and OPLIB opacities \citep{2016ApJ...817..116C}, and the \cite{2009ARA&A..47..481A} solar abundance mixture.  See \cite{2016IAUFM..29B.532G} and references therein for details on the stellar evolution and nonadiabatic pulsation codes used for these studies.

We examined the difference in evolution track in the Hertzsprung-Russell diagram, and the calculated frequencies and growth rates of degree $l$=1 $p$ modes for T$_{\rm eff}$ = 7600 K models using OPAL vs.~OPLIB opacities.  The 2 M$_\odot$ model using OPAL opacities evolves at slightly higher luminosity for the same effective temperature. We found slightly higher growth rates using the OPLIB opacities.  

\begin{figure}
\center
    \includegraphics[width=0.4\textwidth]{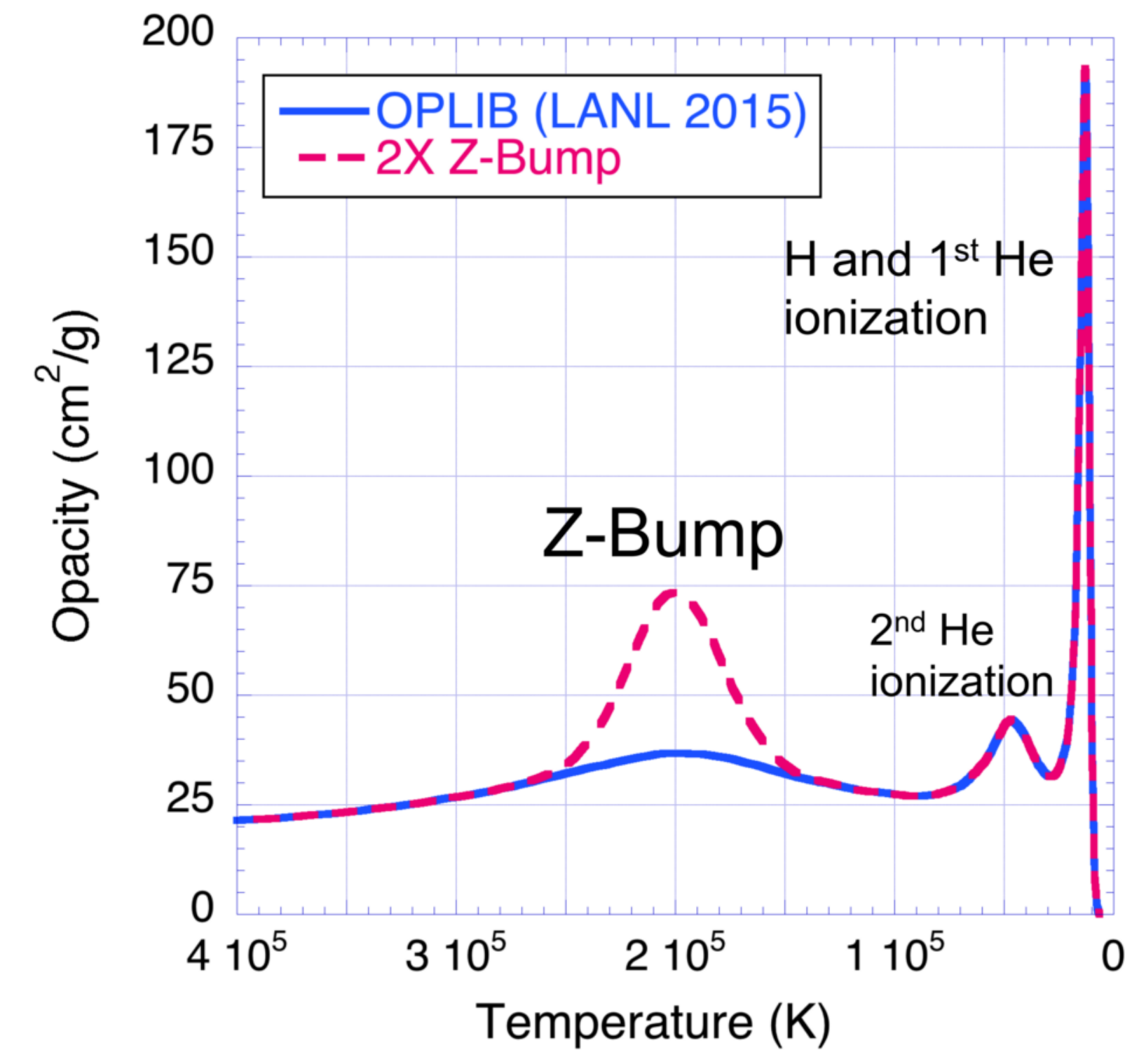}
     \includegraphics[width=0.4\textwidth]{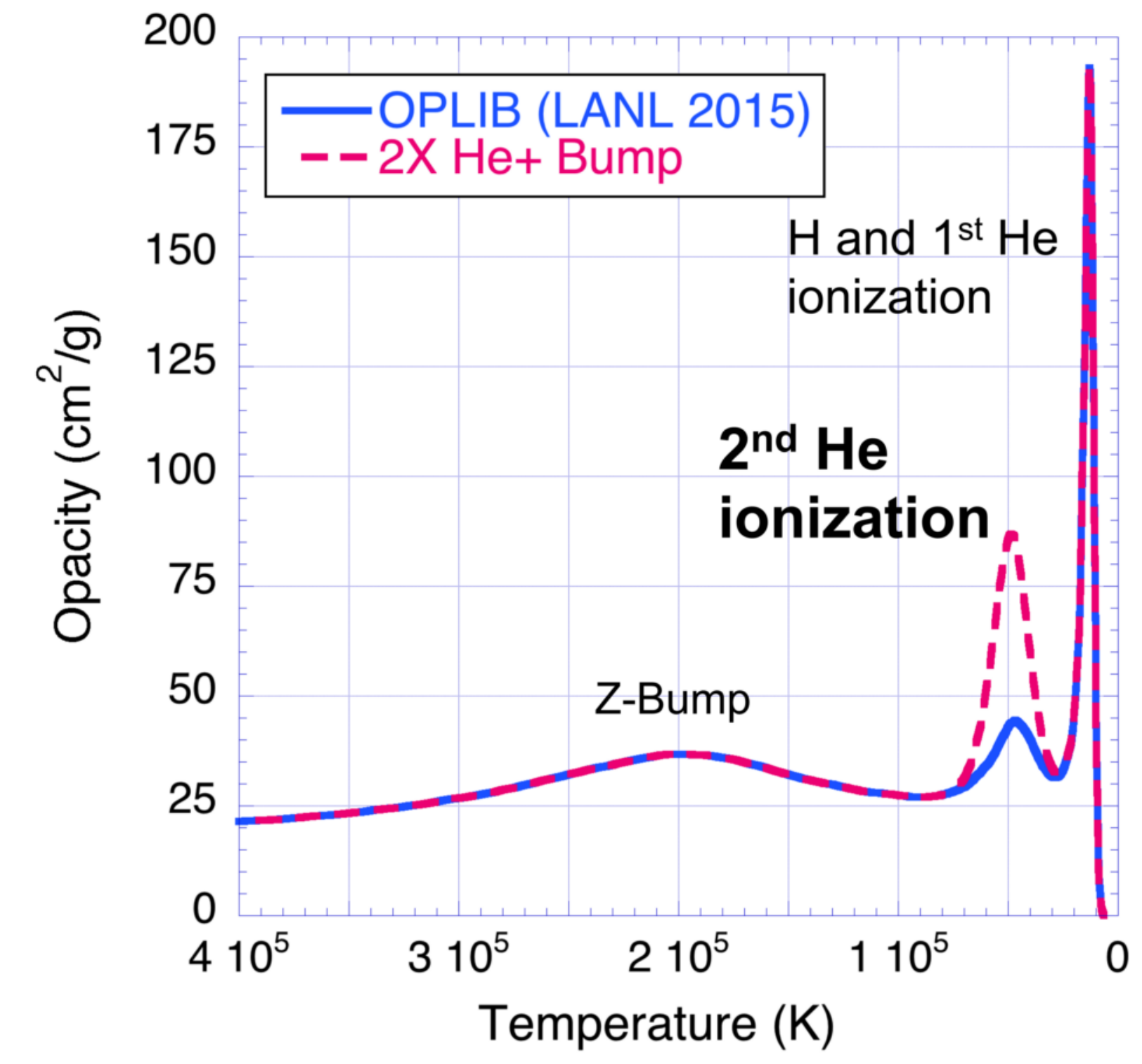}
    \caption{Opacity vs. temperature for 2 M$_\odot$ models with T$_{\rm eff}$  = 7600 K, showing modification of the Z-bump at 200,000 K (left) and of the 2nd He ionization region at 50,000 K (right).}
    \label{fig:Opacity2xZbump}
\end{figure}

\begin{figure}
\center
    \includegraphics[width=0.4\textwidth]{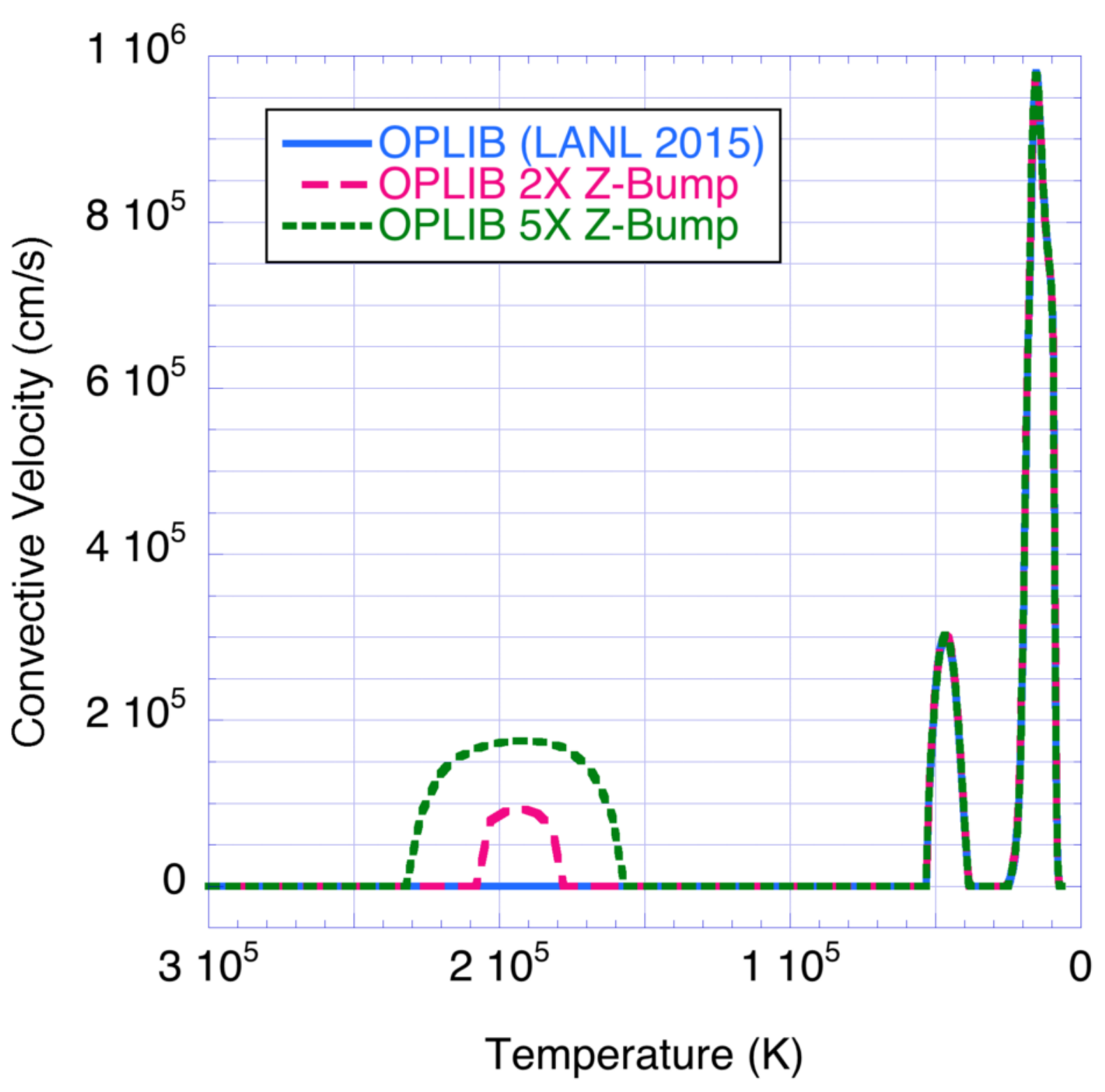}
    \includegraphics[width=0.4\textwidth]{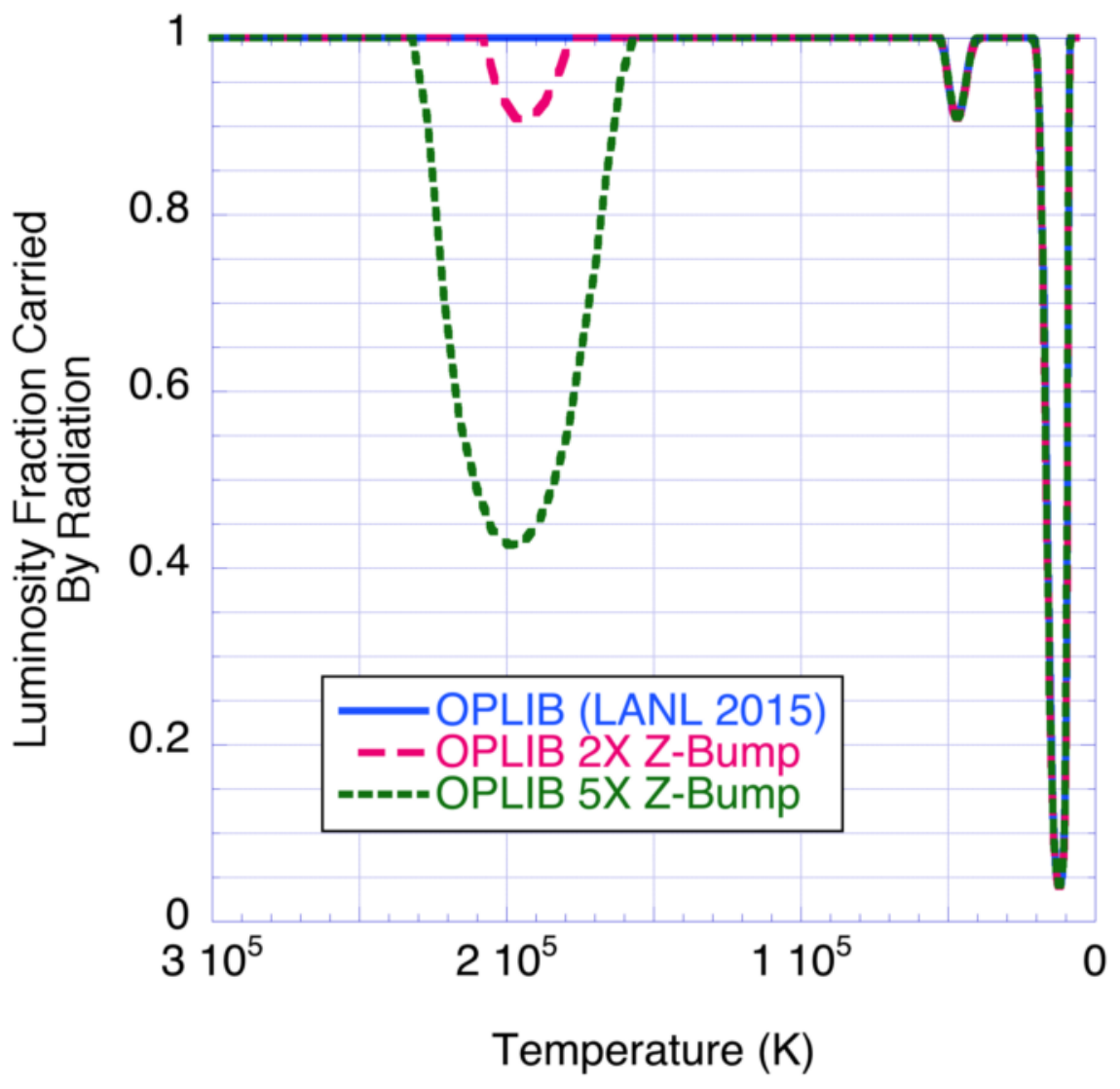}
    \caption{Convective velocity (left) and fraction of luminosity carried by radiation (right) vs.~temperature for 2 M$_\odot$ models with T$_{\rm eff}$  = 7600 K,  with and without $\times$2  and $\times$5 opacity increases in the Z-bump region.}
    \label{fig:CVLFZbump}
\end{figure}

What happens to the evolution and $p$-mode pulsations when Z-bump opacities are increased?  Note that the $\kappa$ effect in the 2nd He ionization region $\sim$50,000 K drives $p$-mode pulsations in $\delta$ Sct stars, so it is not expected that the Z-bump at $\sim$200,000 K will have an effect on the $\delta$ Sct modes.   We modified the opacity by multiplying by a Gaussian function peaking at $\times$2 at 200,000 K (see Fig.  \ref{fig:Opacity2xZbump}).  This Z-bump opacity increase has almost no effect on the 2 M$_\odot$ evolution track.  The Z-bump $\times$2 opacity increase also has almost no effect on $l$=1 $p$-mode pulsations; the growth rates are slightly lower for some modes.   However, the Z-bump opacity increase introduces a small convection zone in 2 M$_\odot$ models near 200,000 K (Fig. \ref{fig:CVLFZbump}).  In the Z-bump region, the convective velocities are 10$^{5}$ cm/s, and convection carries 10\% of the luminosity.

Could neglected broadening cause a spurious need for increased opacity?  Turbulence and convection not taken into account in opacity simulations should broaden lines and features.  It is also possible that photo-excitation processes (Stark line widths) may not be calculated properly in opacity simulations \citep{2016ApJ...824...98K}.  However, \cite{2016ApJ...824...98K} note that a factor of 100 increase in all line widths would be needed to account for missing solar opacity.  The 2nd He ionization opacity region in A-type stars responsible for $\delta$ Sct pulsations is already convective/turbulent.  The lack of evidence for diffusive element settling in A- and F-type stars also indicates that turbulent mixing may be continuously counteracting diffusive settling.

\begin{figure}
\center
    \includegraphics[width=0.4\textwidth]{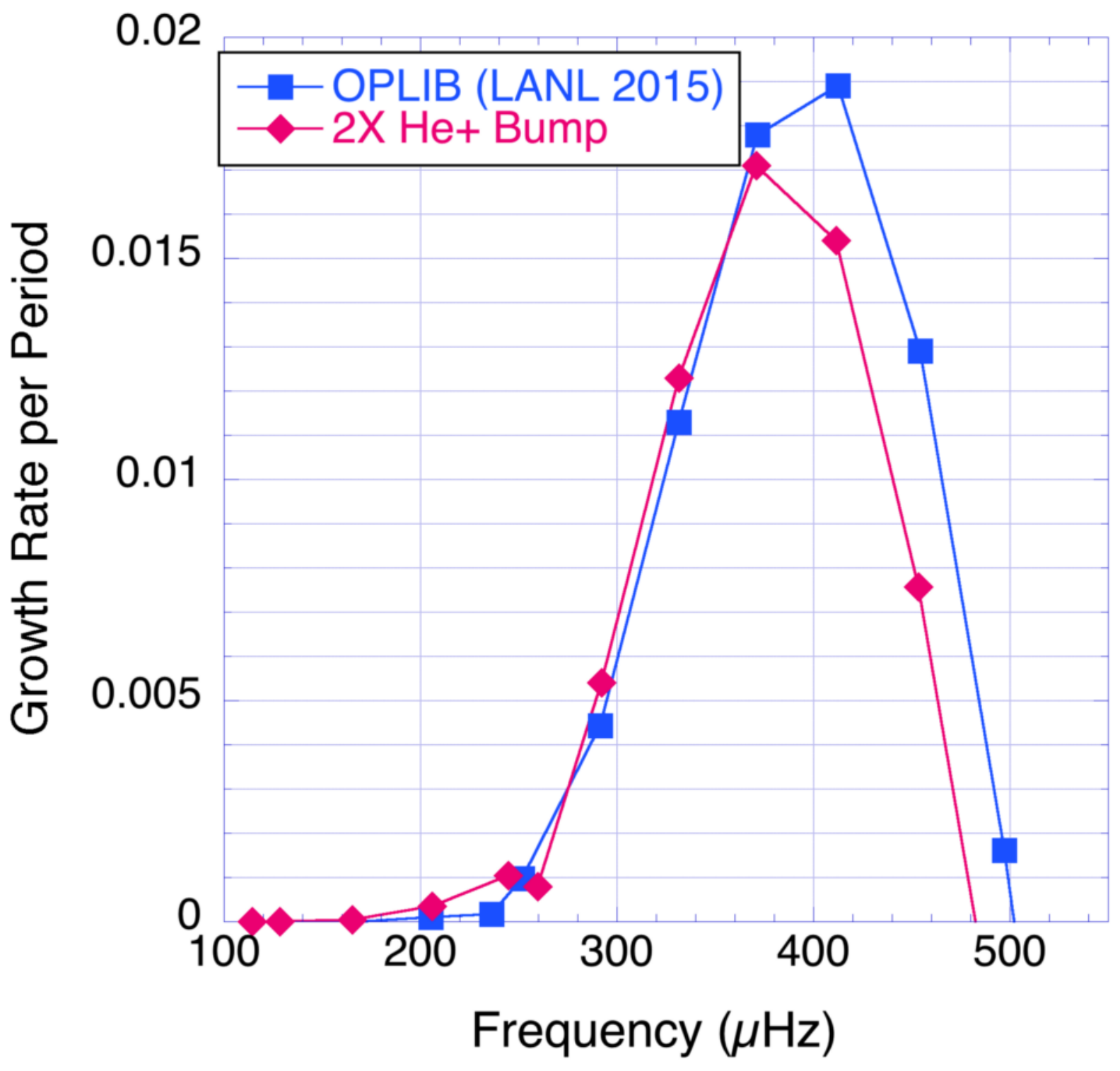}
     \includegraphics[width=0.4\textwidth]{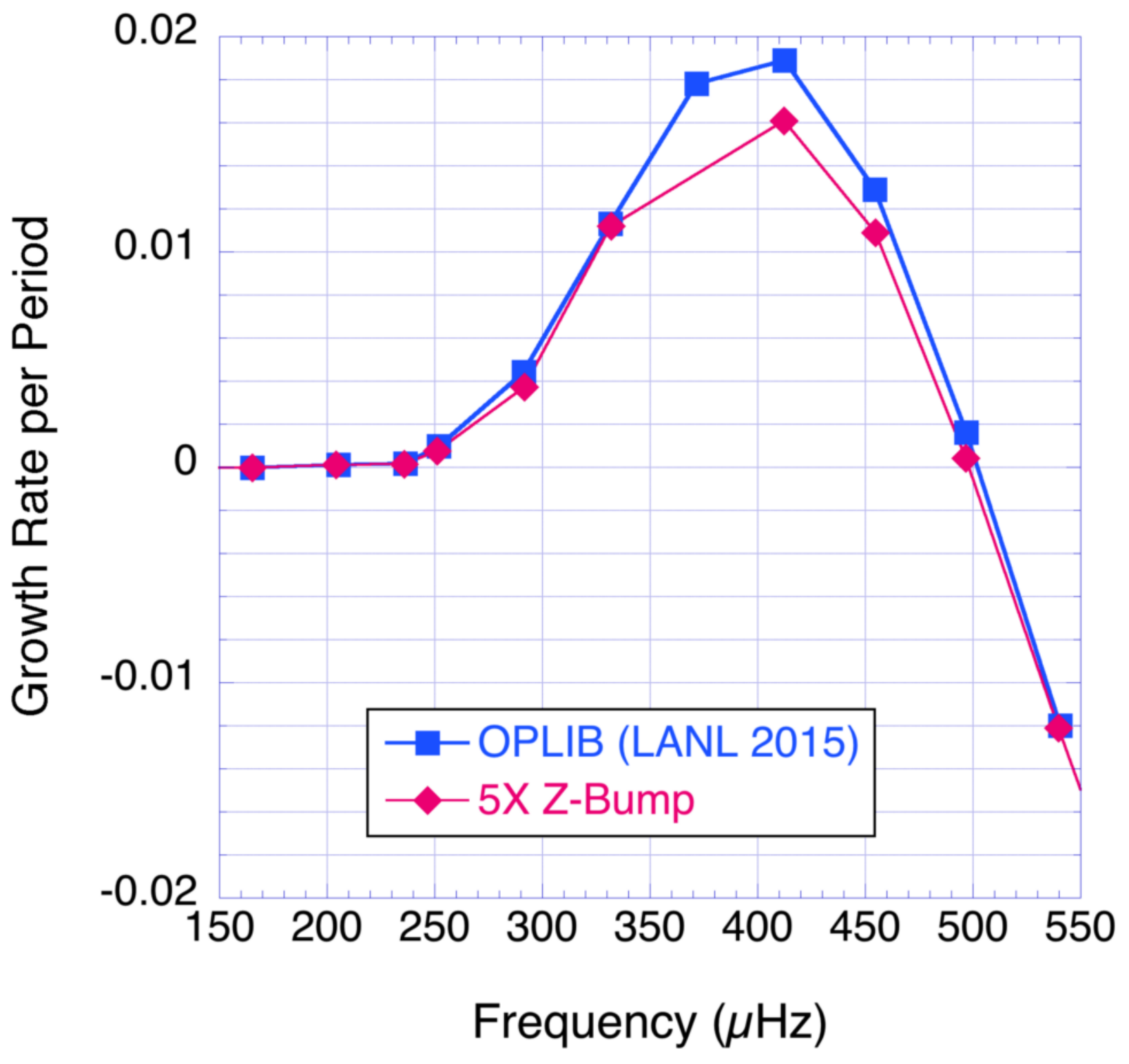}
    \caption{Fractional kinetic-energy growth rate per period (dimensionless) vs.~frequency for $l$=1 $p$ modes of 2 M$_\odot$ T$_{\rm eff}$  = 7600 K models with and without opacity enhancements in the 2nd He ionization region (left) or Z-bump region (right).}
    \label{fig:Growth}
\end{figure}

\begin{figure}
\center
    \includegraphics[width=0.4\textwidth]{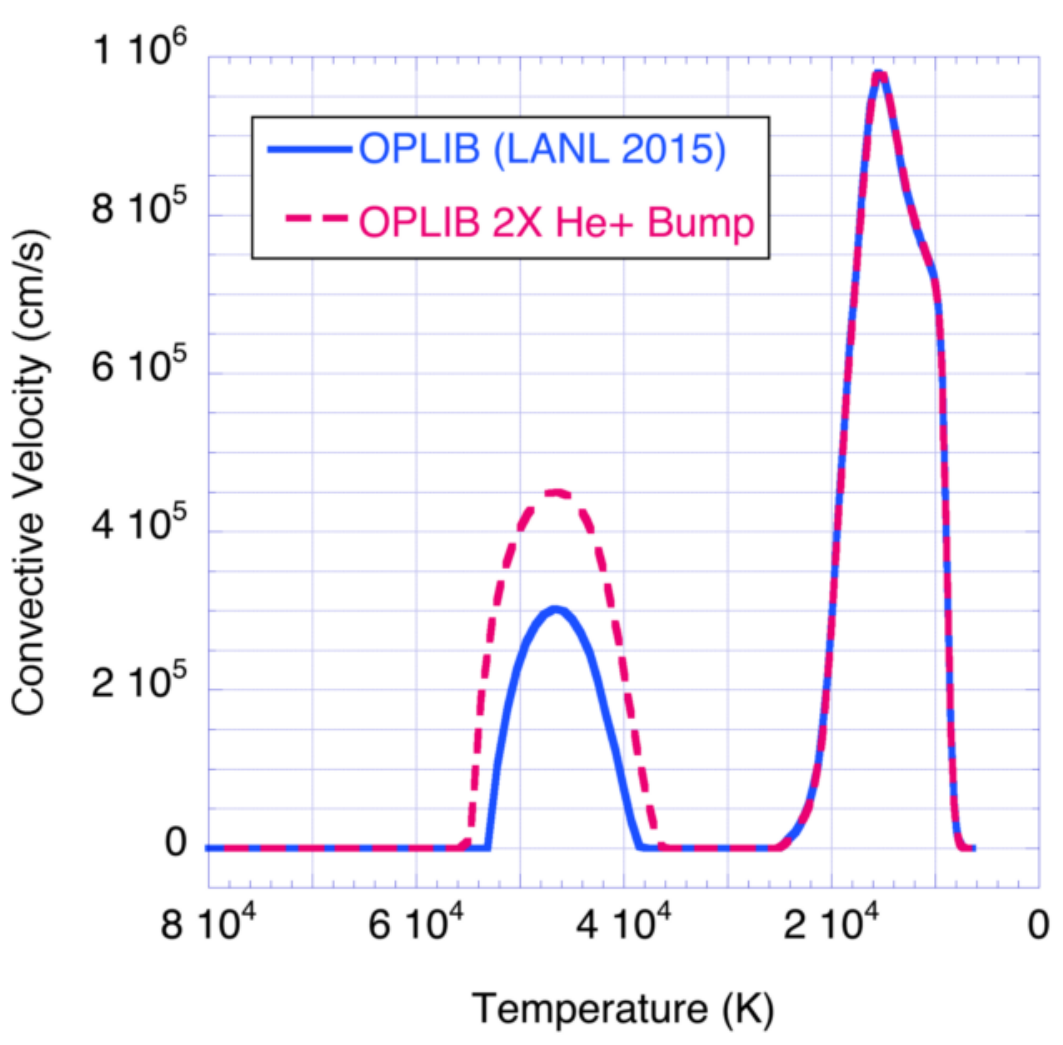}
     \includegraphics[width=0.4\textwidth]{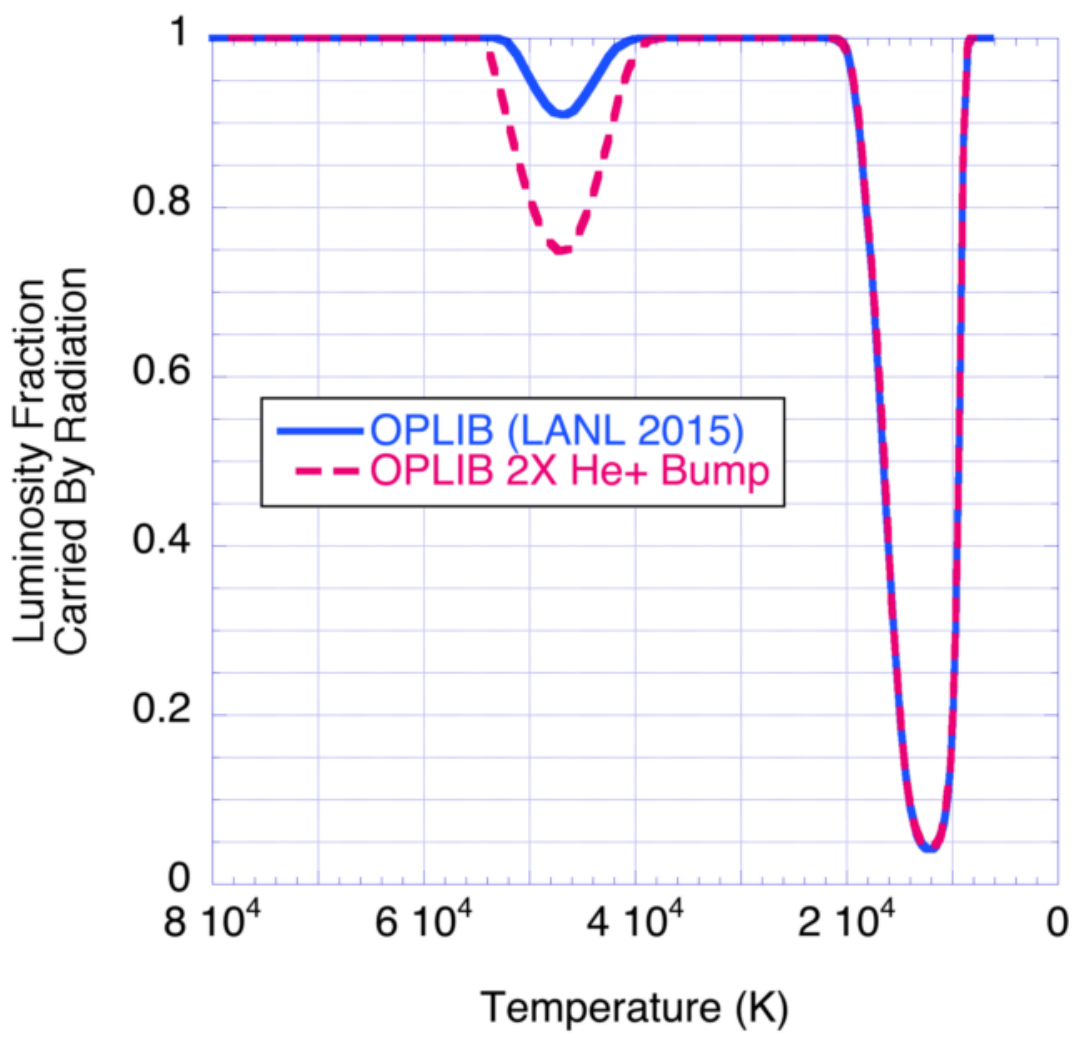}
    \caption{Convective velocity (left) and fraction of luminosity carried by radiation (right) vs.~temperature for 2 M$_\odot$ models with T$_{\rm eff}$  = 7600 K,  with and without $\times$2 opacity increase in the 2nd He ionization region.}
    \label{fig:CVLFHebump}
\end{figure}

What are the consequences for pulsations of A-type stars if turbulence/convection causes an opacity increase in the 2nd He ionization region where $\delta$ Sct $p$ modes are driven?  This opacity bump is produced by ionization, rather than by line excitation; therefore the opacity modification may be more correctly referred to as `edge broadening' or `edge blending'.  We multiplied the opacity by a Gaussian function peaking at $\times$2 in the 2nd He ionization region centered at 50,000 K (Fig. \ref{fig:Opacity2xZbump}).  We find that this He+ ionization region opacity increase has very little effect on the 2 M$_\odot$ evolution track. Increased He+ opacity shifts the unstable $l$=1 $p$-mode frequency range to lower frequencies and also reduces the growth rates (Fig. \ref{fig:Growth}).  Increased He+ opacity causes more luminosity to be carried by convection, weakening the $\kappa$-effect pulsation driving.  With the opacity increase, convective velocities become larger by a factor of two, as high as 4$\times$10$^{5}$ cm/s. Convection then carries 25\% of the luminosity, instead of 10\% (Fig. \ref{fig:CVLFHebump}).

If enhanced opacities cause a small convection zone to develop in the Z-bump region for A-type stars, convection could also cause line broadening that would further increase opacities.  The combination of diffusive settling and radiative levitation can concentrate Fe in the envelope at $\sim$200,000 K, and may also cause a Z-bump convection zone to develop for A-type stars \citep[see, e.g.,][]{2000A&A...360..603T,2012A&A...546A.100T}.

We multiplied the opacities by a Gaussian function peaking at $\times$5 for the Z-bump region centered at 200,000 K.  The  $\times$5 Z-bump opacity increase has almost no effect on the 2 M$_\odot$ evolution track.  However, the $\times$5 Z-bump opacity increase induces a wide convection zone at $\sim$200,000 K, and convective velocities  $\sim$2 $\times$10$^{5}$ cm/s, with convection carrying almost 60\% of the luminosity (Fig. \ref{fig:CVLFZbump}).  The Z-bump $\times$5 opacity increase has almost no effect on $l$=1 $p$-mode pulsations; growth rates become slightly lower for some modes (Fig. \ref{fig:Growth}).  We were hoping that the convection zone might induce $g$-mode pulsations via the convective blocking mechanism proposed for $\gamma$ Dor variables \citep{2000ApJ...542L..57G}.  For T$_{\rm eff}$ = 7600 K models, the $\times$5 Z-bump enhancement does not affect low-order $l$=1 or 2 $g$-mode stability; the low-order $g$ modes are slightly less stable, but are still stable.  One or two of the highest frequency $l$=2 $g$ modes become unstable for cooler (T$_{\rm eff}$  = 7245 K) models, even without an opacity enhancement.  The growth rates become lower with the $\times$5 Z-bump increase, possibly because these modes are driven by the $\kappa$ effect (and not the convective-blocking mechanism) that is weakened by convection.

If turbulence could broaden lines responsible for the Z-bump, and also increase He+ opacity, could turbulence also increase opacity in the H and 1st He ionization region and affect the properties of this near-surface convective region?  In this region there is a high convective velocity, $\sim$2 $\times$ 10$^{6}$ cm/s, and convection carries 90\% of the luminosity (Fig. \ref{fig:CVLFHebump}).  Turbulent pressure (and its perturbation) in the H-ionization zone is proposed as a mechanism to drive coherent modes in A-type stars \citep{2014ApJ...796..118A}, as seen in, e.g., HD 187547 observed by the {\it Kepler} spacecraft.

\section{BRITE observations}

In December 2016, we proposed two A-type stars for BRITE observations.  The first star, HR 7284, with V=6.18, has the same A3V spectral type as HD 187547, but is brighter.  Some short-cadence {\it Kepler} data exist for this star, showing $\delta$ Sct modes, and also possibly modes driven by the turbulent-pressure mechanism.  BRITE observations of this star were judged to be technically feasible by the BRITE-Constellation Executive Science Team (BEST).   The second proposed star is HR 7920 ($\eta$ Ind), a southern hemisphere $\gamma$ Dor/$\delta$ Sct hybrid candidate \citep{2017MNRAS.466..122K}, with V=4.51 and spectral type A9 IV.  Observations of this star were identified by BEST as ``a high priority program that could drive the decision on future fields of observations.''  Neither target has been scheduled yet for BRITE observations.  

\section{Conclusions}

A-type stars may turn out to be useful testbeds for opacity studies.  Convective velocities are 10$^{5}$-10$^{6}$ cm/s and may need to be taken into account in opacity calculations, as they may be responsible for line and edge broadening.  Opacity increases could cause convection zones to appear or widen, abundance gradients to be altered, instability strip boundaries to change, unstable mode frequency ranges to change, and affect operation of driving mechanisms such as the $\kappa$ effect or the proposed driving from turbulent-pressure perturbations.  A-type stars are numerous and bright enough for BRITE observations and follow-up studies.

\acknowledgements{We thank the Department of Energy for funding under the High Energy Density Physics Impact program to attend this conference.}

\bibliographystyle{ptapap}          %RS: the correct one
\bibliography{Guzik}

\end{document}